# Electron-induced chemistry and sputtering of volatile species from amorphous and crystalline water ice

Yogeshwar Nath Mishra[1, ‡, *], Sankhabrata Chandra[1], Murthy S. Gudipati[1, *],
Bryana Henderson[1], Dag Hanstorp[2] and Yuk L. Yung[3]

[1]Jet Propulsion Laboratory, California Institute of Technology, Pasadena, CA 91109, USA
[2]Department of Physics, University of Gothenburg, Gothenburg, SE 41296, Sweden
[3]Division of Geological and Planetary Science, California Institute of Technology, Pasadena, CA 91125, USA

[‡]Currently with Computer, Electrical and Mathematical Sciences and Engineering Division, KAUST, Saudi Arabia
[*]Authors to whom correspondence should be addressed to: Yogeshwar Nath Mishra (mishra@caltech.edu) and Murthy S. Gudipati (Murthy.gudipati@jpl.nasa.gov)

**Abstract**
Electron irradiation of water-rich ices plays a significant role in initiating the chemical and physical processes on the surface of airless icy bodies in radiation environments such as Europa and Enceladus, as well as on the Moon, comets, and asteroids interacting with the solar wind. The sputtering process by electrons and ions leads to chemical modification and outgassing of their icy surfaces and the subsequent formation of a tenuous atmosphere. Though electron-sputtering yields are known to be lower compared to ion-sputtering yields, one needs to also account for their differential fluxes. Europa receives about 10 times more electron flux than proton flux, making electron-induced outgassing important to study and understand. Here, we report a rigorous experimental study of electron bombardment of Europa analog crystalline and amorphous pure water ices at 100 K under ultrahigh vacuum and compare our results with reported data in the literature to provide quantitative data for low-energy electron-induced chemistry and sputtering of water-ice. In our experiments, the electron-induced sputtering yields of all the gaseous species $H_2$, O, OH, $H_2O$, and $O_2$ are investigated for electron energies lower than 2 keV in terms of partial pressure vs. time of irradiation. The effective averaged change in partial pressure of the desorbed species is converted to the number of sputtered atoms or molecules per second per $cm^2$ from the ice, and then to the sputtering yields (number of species sputtered per electron). These values at different electron fluences (electrons/$cm^2$) are then used to derive the number of atoms or molecules desorbed or sputtered per keV of incident electrons, to enable utilizing our data to quantify the contribution of electron-sputtering process towards the exospheric composition of Europa. Our data agrees well with the previously reported data for the sputtering of $O_2$ and $H_2O$ yield for the amorphous ice. We also find that crystalline ice shows significantly lower sputtering yields when compared to amorphous ice, in agreement with the observation of similar trends in the literature. Our work shows that sputtering yields per keV of O, OH, $O_2$, and $H_2O$ drop with increasing electron energy from 0.5 keV to 2 keV. Though we focused on Europa's surface, our results apply to any icy surface receiving solar wind electrons or magnetospheric electrons of outer solar system icy bodies, including the moons of Uranus and Neptune.

# 1. Introduction

The surfaces of planetary icy bodies in our solar system are continuously bombarded with electrons, ions, and photons from the Sun in the form of the solar wind (Neugebauer, 1990) and solar photons (Cooper et al., 2001; R.E. Johnson, 2004) or from the planetary magnetosphere, for example, Jovian's moons Europa, Ganymede, and Callisto receive significant flux of electrons between a few keV to a few hundred MeV, their flux decreasing with increasing energy of electrons, and ions with similar energies and fluxes that are an order of magnitude lower than electrons, due to Jupiter's magnetosphere (Paranicas et al., 2007). Saturnian moons also receive >1 MeV electron radiation (Kollmann et al., 2018). Uranian and Neptunian magnetospheric radiation also bombard their respective moons (Weiss et al., 2021). This radiation, whether solar wind or local magnetospheric, causes surface ice chemistry and results in sputtering/outgassing of volatiles from the icy surface – resulting in the formation of a thin atmosphere around these bodies, also known as exosphere (Smyth and Marconi, 2006). While proton (Shi et al., 1995) and ion-sputtering (Cassidy et al., 2013; Johnson and Sundqvist, 2018) efficiencies are higher compared to electron-induced sputtering (Galli et al., 2018; Heide, 1984; Meier and Loeffler, 2020), ejecting both volatile species and ice grains, fluxes of electrons reaching Europa are higher (Cooper et al., 2001; Johnson and Sundqvist, 2018), making their role significant in outgassing of volatiles, thus contributing to the surface-atmosphere exchange. In contrast to ions, electrons (at a given energy) penetrate by far the deepest into the ice and initiate chemical reactions (Leonard et al., 2018; Nordheim et al., 2018; Paranicas et al., 2009). Some of the primary and secondary dissociation and ionization products such as H, $H_2$, O, and $O_2$ are mobile at the ambient surface temperatures of Europa, while others such as OH, and $H_2O_2$ may be less mobile at 100 K. With increasing ion and electron energy, the production of these species reaches larger depths from the surface, from several micrometers to centimeters.

Over the past few decades, several laboratory studies have been conducted to mimic the surfaces, chemical compositions, and temperatures of icy bodies using ions (Brown et al., 1980; Shi et al., 1995) and electron irradiation (Abdulgalil et al., 2017; Barnett et al., 2012; Floyd and Prince, 1972; Orlando and Sieger, 2003; Sieger et al., 1998) of water ice, compared laboratory data to observations (Berdis et al., 2020), and some presented quantitative electron-induced desorption of water and its radiolysis products (Galli et al., 2018; Meier and Loeffler, 2020; Zheng et al., 2006a). Reliable laboratory data by ion and electron irradiation can be used to estimate the expected exospheric $O_2$ contribution from water ice on Europa (Cassidy et al., 2013), Dione and Rhea (Teolis et al., 2010; Teolis and Waite, 2016). However, ion-induced sputtering is different from the electron-induced sputtering. Ions make quasi-elastic collisions with the surface molecules and sputter them directly from the surface. Irradiation with electrons on the surface can initiate dissociation and ionization of molecules to result in energetic atoms (such as O), molecules (such as $O_2$), and radicals (such as OH). The excess energy deposited in these radiation-chemistry products facilitates their desorption from the icy surface, resulting in the sputtering of chemically altered products into the exosphere. High-energy electrons also generate secondary electrons at lower energies in the ice as a result of the ionization of atoms and molecules. These secondary electrons cause further chemistry at greater depths in the ice.

The ejection of atoms/molecules from the surface is often referred to in the literature as desorption or sputtering. Desorption is quantified as number of molecules/sec or the number of molecules/cm$^2$ (Zheng et al., 2006a; Zheng et al., 2006b; Zheng et al., 2007), while the sputtering yield is defined as the number of molecules ejected per incident projectile i.e. electrons, ions, photons, etc. (Galli et al., 2016; Galli et al., 2018; Galli et al., 2017). Here, we aim to compare relevant literature on the topic and conduct new laboratory experiments to quantify these parameters for crystalline and amorphous ice.

Several research groups have determined the desorption or sputtering from $H_2O$ ice. For example, Zheng et al. have studied the desorption of gases from pure $H_2O$ ices (Zheng et al., 2006a) and pure $D_2O$ ices (Zheng et al., 2007) both at 12 K, and have studied their temperature dependence (Zheng et al., 2006b). The desorption of $H_2$, $O_2$, and $H_2O_2$ were $1.43 \times 10^{16}$ molecules/cm$^2$, $>5.84 \times 10^{14}$ molecules/cm$^2$, $>9.17 \times 10^{13}$ molecules/cm$^2$, respectively, for electrons of an energy of 5 keV and currents of 10 µA (Zheng et al., 2006a). The desorption rates of $D_2$, $O_2$, and $D_2O_2$ in amorphous ice were found to be 32%, 91%, and 7% higher than crystalline $D_2O$ ices (Zheng et al., 2007) at 10 µA and with 5 keV electrons. In recent years, Galli et al. (Galli et al., 2016; Galli et al., 2018; Galli et al., 2017) have reported laboratory studies on sputtering yield of water ice along with some of the issues associated with their quantification. For example, surface charging of thick porous ice water ice layers irradiated with electrons was reported for energies >2 keV in (Galli et al., 2016). Amorphous water-ice films at 91-93 K were sputtered with electrons of energy 0.1-3 keV and fluxes of 6-60×10$^{12}$ electrons s$^{-1}$ cm$^{-2}$ (Galli et al., 2017) and sputtering yields were measured by the loss rate of water-ice films using cryogenic quartz crystal microbalance. These studies reported sputtering of $H_2$, $O_2$, and $H_2O$ as ejected species. For ice at 91 K, the sputtering yields were $Y_{H2O}=(0.1\pm0.1)Y$, $Y_{H2}=(0.6\pm0.1)Y$, and $Y_{O2}=(0.3\pm0.1)Y$, where Y is the energy-dependent yield per incident electron and is equal to (2.3±0.8) for 3 keV electrons. Teolis et al. (Teolis et al., 2017) have presented a detailed review and provided a quantitative model of water ice radiolytic yields of $O_2$, $H_2$, and $H_2O_2$ for any projectiles, energy, or temperature. However, the equation used in the model (Teolis et al., 2017) made a factor of three times underestimation of the yields reported by (Galli et al., 2017) for 3 keV electron irradiation.

In another recent study, Meier and Loeffler reported the sputtering yield of water ice at 60 K by electrons of 0.5-10 keV using microbalance gravimetry (Meier and Loeffler, 2020). When compared to the model in (Teolis et al., 2017), the measured trends with electron energy agreed reasonably with the model, but, the absolute yield was overestimated by at least a factor of three by the model. In another recent work, the contribution of electrons to sputter-produced $O_2$ exosphere on Europa was studied (Davis et al., 2021). From the experiments on the sputtering yield and ejected $O_2$ flux as a function of temperature and an electron energy of 0.5 keV, they concluded that the global surface production rate of $O_2$ from electron-induced sputtering is larger than the production rate previously estimated for all the ion-induced sputtering yields combined. Another recent study investigated energy and temperature-dependent sputtering of ice films and compared the previous literature work to reevaluate the sputtering yield (Carmack and Loeffler, 2024) which shows that the electron sputtered contribution to exospheric $O_2$ is significant in high flux environments like Europa. Based on the literature presented above, we conclude that electron-sputtering of icy surfaces is a key contributor to exospheric volatiles and that their sputtering yields vary from experiment to experiment and need to be better constrained. Further, sputtering of oxygen atoms, which play an important role in charge-exchange with Jovian magnetospheric $O^+$, need to be quantified as well.

In this article, we present the first comprehensive experimental study that quantifies desorption of $H_2$ (m/z=2), O (m/z=16), OH (m/z=17), $H_2O$ (m/z=18), and $O_2$ (m/z=32) as a function of electron fluence and energy for both amorphous and crystalline water ice forms at 100 K. We quantified sputtered volatiles through a "step-wise" temporal variation in their partial pressure when bombarded with electrons at different fluxes at a given energy and repeated these experiments at different energies 0.5 keV, 1 keV, and 2 keV (e.g., Figure 1, 1-5 µA at 2 keV). We also compared the desorption of $O_2$ deduced in the experiments with a chemical-transport model, which has been designed to realistically simulate chemical processes occurring in the ice during electron irradiation (Li et al., 2020). While a qualitative study on the production of O atoms in $C_6H_6$-water ice under electron irradiation was reported earlier (Abdulgalil et al., 2017), our work reported here is the first to quantitatively determine production yields of O atoms along with other sputtered species from both crystalline and amorphous ices. We also report significant differences in the sputtering yields of volatile species between amorphous and crystalline ices, which are compared with the existing literature data. Our work provides necessary quantitative data that can be used for modeling observational data on electron-induced chemistry and sputtering from pure water ice with keV electrons.

## 2. Experimental Setup and Procedures

*2.1 Experimental Details*

Figure 1 shows the schematic of the experimental setup and chemical species detection scheme to investigate the sputtering of pure water ices by electrons. Figure 1(a) depicts the top-view of the experimental setup which consists of an octagonal cryogenic ice vacuum chamber (Kimball Physics, Inc., USA), a sapphire ice sample holder (Viewport, Kurt J Lesker Company, Inc., USA), an electron gun (Model: ELG-2/EGPS-1022, Kimball Physics, Inc., USA) and a mass spectrometer (RGA300 with electron multiplier, Stanford Research Systems, Inc., USA). The pressure of the ice chamber was monitored using a pressure sensor. All experiments were performed on ice under ultrahigh vacuum better than $1 \times 10^{-8}$ mbar as measured by the pressure sensor and the major rest gases in the chamber are $H_2$ and $N_2$. The ultrahigh vacuum environment is crucial for the sputtering experiments and during sputtering the partial pressure of $H_2O$ which is also major in the residual gases, is close to $10^{-10}$ mbar. Therefore, during sputtering experiments at 100 K, the surface will be less contaminated by the residual gases. The sample holder was mounted on a closed-cycle helium cryostat (DE-204SB Cryocooler, Advanced Research Systems, Inc., USA). The temperature of the ice chamber was controlled using a programmable cryogenic temperature sensor 5-500 K (DT-670 silicon diode, LakeShore Cryotronics, Inc., USA). The ice films were vapor deposited on the sapphire window using the degassed ultrapure water (JT Baker Chemicals, Inc., USA) at $\sim 5\times10^{-7}$ mbar chamber pressure through the controlled leak valves and a gas tube onto the sample holder at 150 K (crystalline ice) and

100 K (amorphous ice) for four and two hours, respectively. The sticking coefficient of $H_2O$ on the sapphire window is significantly smaller at 150 K compared to 100 K, and as a result, to obtain similar ice thicknesses that ensure full penetration of electrons through the ice (not substrate), we doubled the deposition time for the 150 K crystalline ice experiments. This procedure yielded ices of about a micron thickness on a 20 mm diameter substrate, with a deposition rate of ~ 8.5 nm per minute for the 100 K experiments and ~4.2 nm per minute for the 150 K experiments (Barnett et al., 2012).

The ice was deposited at normal incidence to the deposition surface. We mark this as the ice deposition position 1 at 0° as shown in Figure 1(a). Other positions when the ice substrate was rotated to conduct electron bombardment experiments are given with reference to this initial deposition position. Initially, for the electron-induced ice desorption studies, the ice sample was rotated to position 2, a 135° clockwise rotation such that the ice directly faced the electron gun. In this initial configuration the quadrupole mass spectrometer (QMS) probe was directly mounted to the chamber, with the ionization region of the QMS be placed inside the vacuum chamber, directly in the view of the electron gun's outlet making a 45° angle, marked as distance "d" in Figure 1(a). Under this configuration, we realized that the electron gun (hence electrons emitted by the gun) was interfering with the QMS strongly resulting in continuum spectra in the mass-scan due to the charge-induced anomalies (see more details in Appendix A1). We rectified this problem by moving the QMS further away at 2d distance using a spacer flange and by rotating the ice sample to position 3, which is 157.5° with respect to the deposition angle. With these adjustments, the QMS ionization region was no longer in the direct view of the electron gun and the anomalies in the analog spectra disappeared. For the rest of the studies reported here, the configuration with position 3 was used. Ice samples were irradiated with electrons at energies of 0.5 keV, 1 keV, and 2 keV, respectively. The electron current was set in the controller unit varying up to a maximum of 5 µA, which corresponds to $10^{13}$ electrons $cm^{-2}$ $s^{-1}$ at the ice target based on our calibration experiments using a Faraday cup in the position of the ice target. The grid voltage and 1st anode voltage were chosen as 1% and 10% of the electron energy, respectively. The focus voltage was fixed to 100 V throughout the experiments. The electron beam diameter was estimated to be 20 mm. We quantified the chemical species leaving the irradiated ice films by measuring their partial pressure variation as a function of time with the mass spectrometer and converted these partial pressures to number densities (see Figure 1(b)). The scan rate of the mass spectrometer was 200 milliseconds per atomic mass unit.

With electron-impact ionization mass spectrometry, fragmentation of molecules to smaller molecular fragments or atomic species is an important process that needs to be accounted for when deriving quantitative sputtering yields of each of these species. We achieved this as follows. To determine the contribution from $H_2O$ fragmentation in the QMS to OH and O mass peaks as well as $O_2$ fragmentation to O mass peaks, we did calibration measurements under similar ice temperatures. For this, we determined the background partial pressure ratios of $O/H_2O$, $OH/H_2O$, and $O/O_2$ during the ice deposition (during two independent experiments, one with $H_2O$ only and the one with $O_2$ only) and found them to be 0.029, 0.27, 0.10, respectively. In comparison, the fragmentation ratios for $O/H_2O$ and $OH/H_2O$ are 0.009 and 0.21 for gas-phase $H_2O$ from the NIST WebBook

(Standards and Technology, 2000), which is somewhat different from what we recorded in our mass spectrometer, potentially due to different instrumentation conditions used. Similarly, the O/O$_2$ ratio from the NIST Web Book is 0.2, which is higher than the value observed in our spectra of 0.1. Though the electron energy is the same, 70 eV, electron current density, or the pumping speed of the vacuum pumps could have been different between our experimental system and the NIST reported experimental system. This would change the residence time of gas-phase molecules in the ionization region, resulting in less post-ionization fragmentation in our experimental conditions. For this reason, we used the ratios derived by our calibration studies under our experimental conditions to improve the quantification and reliability of our analysis.

*2.2. Evaluation of the experimental data*

Figure 1(b) illustrates the experimental data showing "ladder-shaped" partial pressure vs. time plots of ice desorption profiles at constant electron energy and varying electron current. In Figure 2 and Figure 3, the electron current was first increased upwards from 0 to 5 µA and then decreased downwards from 5 to 0 µA each with a step-size of 10 minutes (600 seconds). By studying both directions, we aimed to understand the overall "memory effect" of continuous irradiation. For an electron energy of 2 keV, the partial pressures (P) of H$_2$, O, OH, H$_2$O, and O$_2$ changed as a function of electron current. From the desorption profiles, the mean partial pressure change (ΔP) was calculated for 0-5 µA upward steps and then for 5-0 µA downward steps for the middle 200 seconds (see Figure 1 (b) and Appendix A2). The summation of each resulting ΔP is the effective averaged change in partial pressure, $\Delta P_e$, which was converted to the number of molecules-atoms/cm$^2$, $N$, according to (Zheng et al., 2006a; Zheng et al., 2007) through the following equation:

$$N = \left(\frac{\Delta P_e \cdot S_{eff} \cdot A_v}{RT}\right) \cdot t \qquad (1)$$

where $S_{eff}$ is the effective pumping speed, 162 liters/s for H$_2$ and 193 liters/s for the rest of the gaseous species pumped from the cryogenic chamber. The effective pumping speed was calculated from the pumping speed of the vacuum pump and the conductance between the pump and the vacuum chamber according to (Zheng et al., 2006a). $A_v$ is Avogadro's constant, $R$ is the molar gas constant, $T$ = 100 K, is the temperature of the residual gas, and finally, $t$ is the effective averaged integration time. When $N$ is divided by the electron fluence, we get the number of species per electron, which is then normalized to the incident electron energy to determine the number of species per keV as reported in Table 1. The emission currents set at the electron gun controller were between 1 µA and 5 µA, which are used to describe experimental conditions for ease of following. These settings were then calibrated at the ice position using a Faraday cup (Barnett et al., 2012; Galli et al., 2018) and the incident currents at the ice were determined to be 0.33 µA and 1.65 µA, respectively for 1 µA and 5 µA controller readings, which correspond to a total electron fluence of 4.1×10$^{14}$ and 2.1×10$^{15}$ electrons/cm$^2$, respectively, bombarding the ice over a 200-second period. We also converted $\Delta P_e$ to the sputtering yield (Y) according to (Galli et al. 2018) through the following equation:

$$Y = c\left(\frac{\Delta P_e \cdot S_{eff} \cdot q}{k_B \cdot T \cdot i}\right) \quad (2)$$

Where c is the conversion factor which is assumed to be ~ 1, $k_B$ is the Boltzmann constant, $q$ is the charge of an electron and $i$ is the electron current. Other variables are already expressed in Eq. (1). Y has the units of number of species ejected per incident electron. Like equation (1), we finally converted this to sputtering yield per keV by dividing by the electron energy. These results are summarized in Table 2 and discussed in the result section.

## 3. Results

*3.1 P vs. t profiles*

Figure 2 shows the partial pressure of desorbed gaseous $H_2$ (black), O (red), OH (violet), $H_2O$ (blue), and $O_2$ (green) against time and electron current. Here, the crystalline ice at 100 K was irradiated at 0 to 5 to 0 µA (step size: 1 µA every 600 seconds) at constant energies. At 0.5 keV (short-dashed curves), 1 keV (dashed curves), and 2 keV (solid curves), an evident "ladder-wise" variation of *P* vs. *I* can be seen for $H_2$, O, OH, $H_2O$, and $O_2$. O desorption is more significant at 0.5 keV compared to 1 keV at 5 µA. The response of partial pressure to electron fluence was instantaneous in most of the species except for $H_2$ where a delay was seen. It is consistent with earlier studies reported in (Petrik et al., 2006; Zheng et al., 2006a). The release of $H_2$ is expected from the newly formed $H_2$ molecules near the surface of the water ice. Throughout our experiments, the partial pressure for $H_2$ was > $10^{-8}$ mbar. This is mainly due to the well-known fact that it is hard to pump $H_2$ using conventional turbomolecular vacuum pumps. Similar to Figure 2, the *P* vs *t* plots for amorphous ice at 100 K are given in Figure 3.

Different species have different pumping speeds, sticking coefficients, and sputtering efficiencies. Therefore, to determine the most effective time window for measuring the most accurate partial pressure of each of the species, the whole 600-second sputtering time is split into three parts. During the first 200 seconds of sputtering, the sputtered species may not have reached an equilibrium pressure, therefore, the number of species can be underestimated during this time. By the last 200 seconds, the species have likely reached equilibrium pressure, however, because of the different pumping speeds and different sticking coefficients, some species (notably $H_2$) can accumulate in the chamber more than others and hence can be overestimated. To minimize both these effects, we decided to use the middle 200 seconds throughout this work to quantify the mean partial pressures of each species and convert these partial pressures into the number of species and sputtering yield.

*3.2 Desorption rate and fragmentation-corrected sputtering yields*

Figure 4 shows the average number of desorbed species (i.e., atoms or molecules) per keV for different electron energies starting from 0.5 keV to 2 keV for amorphous and crystalline ice. It is seen from the amorphous ice that with increasing electron energy sputtering efficiency decreases and this agrees with the previous studies (Carmack and Loeffler, 2024; Dupuy et al., 2020). However, for crystalline ice, the number of species for O, OH, and $H_2O$ increases and the number of species for $H_2$ and $O_2$ decreases with increasing electron energy. Uncorrected numbers of each species per keV are also shown in Table 1. However, each of these species is also produced by electron-impact ionization of a parent molecule $H_2O$ in the mass-spectrometer, in addition to have been produced in the ice and sputtered directly from the ice by electron irradiation. Therefore, it is necessary to correct such effects to determine the number of each species sputtered directly from the ice. For example, the $H_2O$ mass spectrum shows for each unit intensity at mass 18 m/z ($H_2O^+$), 0.27 intensity at mass 17 m/z ($OH^+$), and 0.0285 intensity at 16 m/z ($O^+$), due to $H_2O$ fragmentation in the mass spectrometer, as derived from our controlled experiments. Similarly, for each $O_2^+$ observed in the mass spectrum, its fragment $O^+$ is seen at 0.1 intensity. When calculating the number of each species of O and OH sputtered directly from the ice, we removed the fragmentation contributions from the parent molecules $H_2O$ and $O_2$. We did not calculate any backward corrections to the number of each species, e.g., the number of $H_2O$ is not corrected for the 0.27 fraction that goes into the $OH^+$ fragment.

In Table 2 we present the averaged sputtering yields per keV for all the sputtered species. Here, the conversion of $\Delta P_e$ to the sputtering yields per keV is calculated from Eq. (2) for both amorphous and crystalline ice at T = 100 K and E = 0.5 to 2 keV for I = 1-5 µA (incident 0.33-1.65 µA or 4.1 to 21×10$^{14}$ electrons cm$^{-2}$). We found comparable values for desorption and sputtering yields calculated using Eq (1) and Eq (2), respectively. Very minor differences between the two calculations are observed since the values of $t$ in Eq. 1, are approximately ± 200 seconds, deduced while averaging the step-wise partial pressure (please see Appendix A2). When comparing the sputtering yields per keV between crystalline (left) and amorphous (right) ice either from Table 1 or Table 2, in general, the yields per keV are higher in amorphous ice for most energies and species. For example, for $O_2$, the amorphous yields per keV are ~ 1.6 times (2 keV), 2.3 times (1 keV), and more than 3 times (0.5 keV) higher than crystalline yields per keV. For $H_2O$, the amorphous yields per keV are 1.2 times (2 keV), 2.6 times (1 keV), and 4.6 times (0.5 keV) larger. For OH, the yields/keV are 1.09, 2.42, 4.9 times higher at 2 keV, 1 keV, and 0.5 keV, respectively. The yields/keV of O are ~1 (2 keV), ~ 1.2 times for 1 keV, and ~ 4.7 times for 0.5 keV. Finally, for $H_2$, the sputtering yields/keV are similar for 1 keV and 2 keV however, it is ~2 times higher at 0.5 keV for amorphous ice compared to crystalline ice under the same conditions. The values shown in Tables 1 & 2 also demonstrate the importance of internal calibration of the experimental equipment to remove the contribution from fragmentation of parent molecules in the mass spectrometer.

Table 1: Averaged uncorrected and corrected (in parentheses) desorption of number of species per keV – calculated from Eq (1). first: Crystalline (red), second: Amorphous (blue).

| E(keV) | $H_2$ | O | OH | $H_2O$ | $O_2$ |
|---|---|---|---|---|---|
| 0.5 | 8.90/18.70 | 0.011/0.052 (0.0043/0.029) | 0.011/0.054 (0.0010/0.006) | 0.037/0.176 (0.049/0.229) | 0.056/0.176 (0.059/0.184) |
| 1 | 8.25/8.08 | 0.019/0.024 (0.014/0.014) | 0.014/0.034 (0.0016/0.001) | 0.045/0.120 (0.059/0.156) | 0.029/0.067 (0.031/0.070) |
| 2 | 7.24/7.19 | 0.031/0.027 (0.027/0.019) | 0.021/0.023 (0.005/0.0034) | 0.059/0.072 (0.077/0.0934) | 0.027/0.045 (0.028/0.047) |

Table 2: Averaged uncorrected and corrected (in parenthesis) sputtering yields per keV – calculated from Eq (2). Left: Crystalline (red), Right: Amorphous (blue).

| E(keV) | $H_2$ | O | OH | $H_2O$ | $O_2$ |
|---|---|---|---|---|---|
| 0.5 | 3.15/6.80 | 0.0036/0.017 (0.0014/0.009) | 0.0036/0.017 (0.00003/0.002) | 0.012/0.058 (0.016/0.075) | 0.018/0.058 (0.019/0.061) |
| 1 | 2.96/2.93 | 0.006/0.0080 (0.004/0.005) | 0.004/0.011 (0.00005/0.0003) | 0.015/0.039 (0.019/0.051) | 0.009/0.022 (0.010/0.023) |
| 2 | 2.64/2.61 | 0.010/0.0087 (0.008/0.007) | 0.007/0.0075 (0.0018/0.0011) | 0.019/0.023 (0.025/0.031) | 0.008/0.0149 (0.009/0.015) |

Figures 5 and 6 show the desorption rate of gases from crystalline and amorphous ices - as the number of molecules-atoms cm$^{-2}$ keV$^{-1}$ against electron fluence (electrons/cm$^2$) at 100 K. The desorption is deduced by using $\Delta P_e$ (see Appendix A2) and other experimental parameters according to Eq. (1). Figures 5 (a, d) for $H_2$, (b, e) for O, and (c, f) for OH show the plots of desorption for the crystalline and amorphous ices, respectively. Similarly, in Figures 6 (a, c) for $H_2O$ and (b, d) for $O_2$ and their corresponding desorption are plotted as solid curves (0-5 µA, upward steps) and dashed curves (5-0 µA, downward steps). In the case of amorphous ice, the desorption of species from the ice irradiated at 0.5 keV is the most efficient. Such a trend, "decreasing sputtering yields with increasing energy" in this electron energy region, has been observed before (Meier and Loeffler, 2020). We attribute this behavior as being due to the penetration depths of electrons in ice. While 0.5 keV electrons dissipate most of their energy on the surface or near-surface (<50 nm), causing molecular dissociations and ionizations that can lead to easier release of molecules and atoms into vacuum from the surface, higher energy electrons penetrate deeper, where these species are produced. Their kinetic energy is dissipated during collisions before reaching the surface. Hence at higher electron energies, radiolysis products end up getting trapped or react with the ice molecules, reducing sputtering efficiency. In the case of crystalline ice, the number of $H_2$ and $O_2$ species sputtered are higher for 0.5 keV and O, OH, and $H_2O$ are higher for 2 keV. This can be explained as crystalline ice is more ordered and has more hydrogen-bonded arrangement, therefore, higher energy is required to sputter. The maximum values of desorption (average for up and down current) for $H_2$, and other gases at the different fluences are given in Table 3 for 2 keV electron energy. Table 3 also provides a direct comparison of the desorption of $H_2$, O, OH, $H_2O$, and $O_2$ for crystalline (left)/ amorphous (right) ices. In general, the production of $H_2$, O, and OH are relatively similar in both crystalline and amorphous ices. Comparing the number of sputtered species for a particular current while upward current vs downward current shows that the number of species from the upward current is lower than the downward current value.

*3.3 Effect of Continued Radiolysis on Sputtering Efficiencies*

We observed an unusual behavior in the case of $H_2O$ and OH (mostly fragmentation from $H_2O$) sputtering at 0.5 keV. The sputtering yields increase non-linearly with electron flux – unlike other energies and species that show more or less a linear increase with electron flux (Figures 5 & 6). While upward increase in the electron flux, the sputtering yields are lower first and increase more with higher electron flux. This could be rationalized by the possibility that the higher the electron-flux, the higher would be the energy deposited per unit area that not only helps break hydrogen bonding in the water-ice, liberating individual molecules, but also provides sufficient kinetic energy for these molecules to escape ice surface. We also observe a "hysteresis" like effect when ramping down the electron flux. For a given electron flux, stepping down from higher flux irradiation produces higher sputtering than stepping up from a lower flux. This is also in line with the above explanation that higher flux would break larger amounts of hydrogen bonds on the surface making these molecules ready to leave the surface. Hence, we propose a two-step process of electron-induced sputtering of $H_2O$ molecules from the water-ice surface.

The first step involves the electron-induced breaking of hydrogen bonds, while the second step is to provide these $H_2O$ molecules with the necessary kinetic energy to leave the surface from the next impinging electron. At higher electron energies (1 keV and 2 keV), we do not see this behavior because, as discussed earlier, the majority of their energy is deposited into greater depths. We observe here the expected linear increase and decrease with changing electron flux in either direction within the error limit (Figures 5 & 6). We also observe in Figure 5d that for crystalline ice, the initial production of $H_2$ is higher at lower fluxes and later the sputtering yields follow the expected linear increase and decrease with changing fluxes of electrons. Though the error bars are higher than the unique deviation from the rest of the data, we could interpret this as due to a few water molecules that are trapped between the grain-boundaries of otherwise compact, highly oriented, crystal lattice of crystalline water-ice. Once these unbound surface water molecules are dissociated, the rest of the ice behaves like crystalline ice for $H_2$ production. We observed similar behavior but peaking at higher flux for 1 keV electrons (Figure 5d) – higher sputtering yields from freshly made ice as the electron flux is increased and returns to normal yields subsequently.

Key insights from the presented results:

1) The sputtering yields, normalized to energy, are higher for 0.5 keV electron bombardment compared to 2 keV, aligning with existing literature. However, our findings indicate that the sputtering of chemical species increases with electron energy from 0.5 keV to 2 keV, likely due to the memory effect induced by continuous electron irradiation.
2) Sputtering yields for amorphous ice are higher compared to crystalline ice at 0.5 keV electron energy.
3) $H_2O$ and its fragment OH show non-linear flux dependent sputtering for 0.5 keV electron bombardment.
4) $H_2$ production is higher from freshly made ice with 0.5 keV and 1 keV electrons.

**Table 3:** Number of species/cm² desorbed (uncorrected and corrected in the parentheses) vs. incident electron fluence for $H_2$, O, OH, $H_2O$, and $O_2$ (Left: Crystalline (in red), Right (in blue): Amorphous) at electron energy 2 keV and current 1-5 µA corresponding to electron fluences of 4.1 to 20.7×10¹⁴ electrons cm⁻².

| Incident fluence ($\times 10^{14}$) | $H_2$ ($\times 10^{16}$) | O ($\times 10^{14}$) | OH ($\times 10^{14}$) | $H_2O$ ($\times 10^{14}$) | $O_2$ ($\times 10^{14}$) |
|---|---|---|---|---|---|
| 4.1  | 0.63/0.62 | 0.26/0.23 (0.23/0.18)  | 0.18/0.20 (0.048/0.023) | 0.50/0.67 (0.65/0.88)  | 0.16/0.35 (0.17/0.37) |
| 8.2  | 1.23/1.23 | 0.52/0.46 (0.45/0.34)  | 0.36/0.39 (0.09/0.058)  | 0.97/1.24 (1.27/1.61)  | 0.42/0.76 (0.44/0.80) |
| 12.4 | 1.76/1.79 | 0.77/0.66 (0.65/0.49)  | 0.53/0.57 (0.013/0.091) | 1.49/1.78 (1.94/2.31)  | 0.71/1.13 (0.75/1.20) |
| 16.5 | 2.35/2.30 | 1.01/0.84 (0.85/0.62)  | 0.70/0.71 (0.17/0.19)   | 1.97/2.21 (2.56/2.87)  | 1.00/1.52 (1.05/1.60) |
| 20.7 | 2.89/2.83 | 1.26/1.03 (1.06/0.76)  | 0.84/0.87 (0.22/0.16)   | 2.33/2.65 (3.03/3.44)  | 1.26/1.90 (1.32/2.00) |

## 4. Discussions

Table 4 presents a summary of our results, comparing the sputtering yields per keV obtained in this study with values reported in the literature. To our knowledge, absolute sputtering yields for all species from a single experiment have not been previously reported. Existing studies focus on specific species, but not all of them in a single study. Therefore, our study provides a comprehensive and quantitative analysis of the sputtering yields for all possible atomic and molecular species from amorphous and crystalline water ices, establishing a common reference study future research. Below, we compare our sputtered yields/keV with the literature values, starting with $H_2$ followed by O, OH, $H_2O$, and $O_2$. The sputtering of $H_2$ was previously reported by Zheng et al. (2006a). The values of sputtering yields for different species are given in Table 4. Please note that the sputtering yield calculated for $H_2$ by Davis et al. (2021) is far from our obtained value. Davis et al. (2021) estimated the yield based on the assumption that the $H_2$ and $O_2$ sputtering occurs stoichiometrically at equilibrium, leading to the assumption that $Y(H_2)=2Y(O_2)$. Our experimental values may be slightly overestimated as $H_2$ is difficult to pump despite accounting for the pumping speed.

Our study is the first to comprehensively report atomic oxygen (O) and hydroxyl (OH) sputtering yields which is an important addition to already available literature data. We found no previous literature on sputtering of O and OH together. For crystalline ice at lower fluencies (Table 3), the sputtering efficiency of O atoms exceeds that of $O_2$. However, at higher fluences, $O_2$ yields become comparable to O yields. This trend is likely due to the production of a higher amount of dissociation products per unit volume at higher fluencies, providing a higher probability for the O atoms to recombine forming $O_2$, before outgassing from the ice surface. In contrast, for amorphous ice, $O_2$ yields are consistently higher than O yields across all fluences. Under planetary conditions, where electron fluences are much lower than in the laboratory setting, we propose that O and $O_2$ sputtering yields would be equally significant, with O potentially dominating at lower fluences. When comparing OH sputtering yields to H2O sputtering yield, the OH sputtering yield is 66x lower in crystalline ice as compared to amorphous ice at 0.5 keV. However, the sputtering yield is very similar between the two ice types at 2 keV. Overall, OH sputtering yield is significantly higher in amorphous ice at 0.5 keV but decreases at 2 keV. This trend likely results from reduced trapping of OH or weaker bond networks in amorphous ice, leading to higher sputtering at lower energies, which declines as electron energy increases. Notably, OH desorption closely follows $H_2O$ desorption curves, reinforcing the reliability of quantification.

Compared to the literature, our $H_2O$ sputtering yield is significantly lower than the value reported by Galli et al. (2018). However, our experimental values are ~ 4 to ~ 14 times lower than those of Meier & Loeffler (2020) and Davis et al (2021), respectively. This significant difference arises from differences in detection techniques: while Galli et al. (2018) used a microbalance to measure mass loss, we used mass spectrometry method to determine partial pressure and calculate the sputtering yield. A fraction of sputtered $H_2O$ molecules may have been lost to the vacuum system or to other parts of the cryostat before detection by the mass spectrometer, leading to an underestimation of total $H_2O$ sputtered yield compared to Galli et al.'s results. Our $O_2$ sputtering yields are approximately 10 times lower than those reported by Teolis et al., (2017) and Loeffler et al. (2020), and ~ 6 times lower than those of Davis et al. (2021). Teolis et al. conducted experiments at a higher temperature (130 K) compared to our 100K, while other experimental parameters remained similar. On the contrary, Loeffler et al., and Davis et al., have used significantly lower fluences. The possible deviation between our results and previous studies may arise due to differences in the structure and porosity of amorphous/crystalline ices made in different laboratories. Variations in experimental geometry, deposition and detection angles, and irradiation times can all influence sputtering yields. In typical deposition conditions, the formed amorphous ice has a porosity of ~17% (Dohnalek et al., 2003) which greatly mobilizes the O atoms. Therefore, in our study a small variation in experimental conditions such as ice thickness and angle of deposition could sputter more $O_2$.

An important observation regarding the step-up vs. step-down of our electron current sequence is that some species exhibit a hysteresis-like memory effect, whiles others don't. Notably, $H_2O$ sputtering shows a significant increase during the step-down current sequence at the same electron fluence. This effect can be caused by both radiation-processed ice surfaces with weakly bound molecules on the surface that can easily be sputtered. Additionally, it could be influenced by the slower pumping efficiency of the system due to the adsorption and the delayed release of these molecules on metal surfaces inside the vacuum chamber. To account for this effect and to improve accuracy, we used the mean values of the two data points when calculating the sputtering yields. Furthermore, because a significant fraction of OH species originate from $H_2O$ fragmentation in the mass spectrometer, OH exhibits a sputtering yields profile similar to that of $H_2O$ as a function of fluence (see Figure 5f with Figure 6c). When describing the electron-induced desorption chemistry of water ice, $H_2O$ can go through a unimolecular decomposition via Eq. (3), (4), and (5). Two H atoms can recombine to form molecular hydrogen in Eq. (6). The OH radical can be formed due to the release of OH via Eq. (3) or via the decomposition of $H_2O_2$ into two OH molecules in Eq. (7) (Zheng et al., 2006a; Zheng et al., 2007). Therefore, it could lead to overestimation of OH molecules. However, the desorption of $H_2O_2$ in our experiments was found to be negligible. The production of OH can also be supported by the oxygen abstraction reaction in Eq. (8). The O atoms are generated through Eq. (4) and (5) while two O atoms can recombine to form $O_2$ molecules in Eq. (9). Finally, $O_2$ can also be produced due to the radiation-induced decomposition of hydrogen peroxide, Eq. (10) as well as OH and O recombination to form $HO_2$ and its fragmentation leading to the production of $O_2$, Eq. (11). Our recent numerical modeling work supports that $HO_2$ is an important intermediary in the formation of $O_2$ in ice (Li et al., 2022).

$$H_2O \rightarrow H + OH \quad (3)$$
$$OH \rightarrow H + O \quad (4)$$
$$H_2O \rightarrow H_2 + O \quad (5)$$
$$H + H \rightarrow H_2 \quad (6)$$
$$H_2O_2 \leftrightarrow 2OH \quad (7)$$
$$O + H_2O \rightarrow OH + OH \quad (8)$$
$$O + O \rightarrow O_2 \quad (9)$$
$$H_2O_2 \rightarrow O_2 + H_2 \quad (10)$$
$$OH + O \rightarrow HO_2 \rightarrow O_2 + H \quad (11)$$

**Table 4:** Comparison of quantitative desorption and sputtering yields of this study with literature. Left: literature data (non-italic font), Right: this study (italic font). The amorphous and crystalline ices are denoted as A and C, respectively. The electron impact angle to the normal of the ice surface is denoted as θ with values in subscript (in our study this angle is 67.5 deg). As different publications provide different units of the yields, we compared our data with the literature data in those units.

| Sputtered species | Zheng et al. (2006a)/ *this study* **(number of desorbed molecules)** | Teolis et al. (2017) / *this study* **(sputtering yields/electron)** | Galli et al. (2018) / *this study* **(sputtering yields/electron)** | Meier & Loeffler (2020) / *this study* **(sputtering yields/electron)** | Davis et al. (2021) / *this study* **(sputtering yields/electron)** |
|---|---|---|---|---|---|
| $H_2$ | $1.43 \times 10^{16}$ (C, 5keV, 10μA, 12K) / *$2.89 \times 10^{16}$ (C, 2keV, 5μA, 100K)* | Not reported/*Table 2* | Not reported/*Table 2* | Not reported/*Table 2* | ~ 0.15 (C, 0.5keV, ~ $3 \times 10^{13}$ e-cm$^{-2}$s$^{-1}$, 100K, θ$_{12.5}$) / *4.97 (C, 0.5keV, 4x10$^{14}$ e-cm$^{-2}$s$^{-1}$, 100K, θ$_{67.5}$)* |
| O | Not reported/*Table 2* | Not reported/*Table 2* | Not reported/*Table 2* | Not reported/*Table 2* | Note reported/*Table 2* |
| OH | Not reported/*Table 2* | Not reported/*Table 2* | Not reported/*Table 2* | Not reported/*Table 2* | Note reported/*Table 2* |
| $H_2O$ | Not reported/*Table 2* | Not reported/*Table 2* | 3.3 (C, 0.5keV, < $1 \times 10^{16}$ e-cm$^{-2}$, 92K, ~ θ$_{45}$) / *0.012 (C, 0.5keV, 2.07 ×10$^{15}$ e-cm$^{-2}$, 100K, θ$_{67.5}$)* <br> 2.2 (C, 2keV, $1 \times 10^{16}$ e- cm$^{-2}$, 92K, ~ θ$_{60}$) / *0.015 (C, 2keV, 2.07 ×10$^{15}$ e-cm$^{-2}$, 100K, θ$_{67.5}$)* | 0.07 (C, 2keV, ~ $1.1 \times 10^{14}$ e-cm$^{-2}$s$^{-1}$, 60K, θ$_{12.5}$) / *0.019 (C, 2keV, 4x10$^{14}$ e-cm$^{-2}$s$^{-1}$, 100K, θ$_{67.5}$)* | ~ 0.165 (C, 0.5keV, ~ $3 \times 10^{13}$ e-cm$^{-2}$s$^{-1}$, 100K, θ$_{12.5}$) / *0.012 (C, 0.5keV, 4x10$^{14}$ e-cm$^{-2}$s$^{-1}$, 100K, θ$_{67.5}$)* |
| $O_2$ | > $5.84 \times 10^{14}$ (C, 5keV, 10μA, 12K) / *~ $1.32 \times 10^{14}$ (C, 2keV, 5μA, 100K)* | ~ 0.1 (A-C, 2keV, ~ $1 \times 10^{15}$ e-cm$^{-2}$, > 130K, θ$_{45}$) / *0.01(C, 2keV, 2.07 ×10$^{15}$ e-cm$^{-2}$, 100K, θ$_{67}$)* | Not reported/*Table 2* | 0.06±0.01 (C, 2keV, ~ $3 \times 10^{13}$ e-cm$^{-2}$s$^{-1}$, 60K, θ$_{12.5}$) / *0.006 (C, 2keV, 4x10$^{14}$ e-cm$^{-2}$s$^{-1}$, 100K, θ$_{67.5}$)* | ~ 0.082 (C, 0.5keV, ~ $3 \times 10^{13}$ e-cm$^{-2}$s$^{-1}$, 100K, θ$_{12.5}$) / *0.014(C, 0.5keV, 4x10$^{14}$ e-cm$^{-2}$s$^{-1}$, 100K, θ$_{67.5}$)* |

## 5. Planetary science implications

Europa has a tenuous, $O_2$-dominated atmosphere which is originated from the outgassing of its icy surface, and both magnetospheric ions and electrons contribute to the release of gases from the ice (Cooper et al., 2001; Hall et al., 1995; Johnson et al., 2002; Roth et al., 2016; Smyth and Marconi, 2006). According to Table 1, taking 2 keV electrons as an example, the estimated rate of electron-induced $O_2$ desorption is ~0.028 $O_2$ per keV for crystalline ice and ~0.047 $O_2$ per keV for amorphous ice. Considering a global electron energy flux of ~$6.2 \times 10^{10}$ keV cm$^{-2}$ s$^{-1}$ at the surface of Europa (Cooper et al., 2001), the corresponding $O_2$ flux from the surface would be ~$1.73/2.91 \times 10^9$ cm$^{-2}$s$^{-1}$ for crystalline/amorphous ice. Previous models use an $O_2$ surface flux of ~$1.9 \times 10^9$ cm$^{-2}$s$^{-1}$ (Li et al., 2020), ~$1.5 \times 10^8$ cm$^{-2}$ s$^{-1}$ (Carmack and Loeffler, 2024; Sieger et al., 1998) and ~$1.7 \times 10^9$ cm$^{-2}$s$^{-1}$ (Carmack and Loeffler, 2024; Teolis et al., 2010) to compute the chemical-transport model for Europa's atmosphere, which is in excellent agreement within the errors of our experimentally derived outgassing values of $O_2$ from water-ice. Unlike proton or ion sputtering, whereby the majority of the energy is converted to eject unaltered materials (here $H_2O$ molecules), electron sputtering causes both sputtering and chemistry in $H_2O$ ice, generating O, $O_2$, $H_2O_2$, and OH species that are also sputtered along with $H_2O$. $O_2$ production, on the other hand, requires chemical processes to occur. For this reason, we propose that the majority of $O_2$ in Europa's atmosphere is due to electron-induced ice chemistry. Note that the higher the kinetic energy of an electron, the deeper it will penetrate the ice before completely dissipating its energy, producing O, $O_2$, $H_2O_2$ as well as OH deeper in the ice. Since the electrons impacting Europa have energy as high as hundreds of keV to tens of MeV (Nordheim et al., 2018), it is possible that the rate of electron-induced $O_2$ desorption would decrease with increasing electron energy, as more $O_2$ is then trapped deeper in the ice. Therefore, experiments on high-energy electron irradiation of water ice are needed to determine the contribution of the chemistry and sputtering by Jovian electrons to $O_2$ in Europa's atmosphere. The sputtering yield of $O_2$ is higher than that of O atoms in both crystalline and amorphous ice.

Understanding OH desorption is crucial for studying Europa's surface and exosphere, as sputtering by energetic electrons and ions contributes to the moon's tenuous atmosphere. The release of OH can influence the formation of molecular oxygen and hydrogen, key components in Europa's potential habitability and surface chemistry. In our study, OH desorption very much follows the trends of $H_2O$ desorption even after rigorously controlling for the proportion of OH that is generated from H2O fragmentation in the mass spectrometer. Further, electron-sputtering produced fluxes of H2O are similar to O2 in our experimental study. However, sputtering caused by ions is expected to contribute significantly to H2O production. Based on these results, we conclude that while both electrons and ions contribute to the sputtering of undissociated $H_2O$ into Europa's atmosphere, radiolysis products such as $O_2$ may be dominated by energetic electron bombardment. In addition, our experiments show that considerable amounts of O and OH are also directly released from the ice into the atmosphere due to electron irradiation. More work on sputtering yields at higher electron energies as well as H$^+$, O$^+$, and S$^+$ on Europa's ice analogs are needed to obtain a comprehensive understanding of radiation-mediated chemistry and sputtering to Europa's atmosphere and improved existing models. Our work is also pertinent to better understand electron-stimulated chemistry and desorption from comets and other bodies with surface ice layers. We show that a significant amount of $O_2$ is desorbed with electron-bombardment of water ice. There is a debate on the origin of $O_2$ in cometary outgassing from the Rosetta mission. While some attribute to primordial $O_2$ (Heritier et al., 2018), others contest this (Yao and Giapis, 2018) with the possibility of low-energy ion bombardment of $H_2O$ ice forming $O_2$ at the surface of the comet. Our work shows that in general, in-situ radiolysis with low-energy electrons can also be a source for the formation of $O_2$ in cometary outgassing. Similarly, solar wind electrons could contribute to the sputtering of $H_2O$ and $O_2$ from airless asteroid surface ice such as on Ceres (Raponi et al., 2018), causing transient $O_2$ exospheric mixing. On the Moon, particularly in the permanently shadowed regions (Paige et al., 2010) that are very cold (~30 K) but are exposed to the solar wind, the amorphous ice surface is expected to release $H_2$, O, $O_2$, and OH, of which H loss exceeds O loss (as per our and other data). As a result, these icy surfaces on the Moon may be enriched with $H_2O_2$. Finally, low-energy electron-induced surface ice chemistry could produce oxygen species in interstellar and circumstellar ice grains exposed to local electron irradiation, promoting the O-atom-involved chemical evolution of organic matter. We found the mobility of O atoms in amorphous ice to be higher than in crystalline ice, producing more $O_2$. Interstellar ice grains are mostly amorphous, providing the environment for oxidation of organics as well as $O_2$ production in those ice grains.

## 6. Conclusions

An experimental investigation was performed concerning electron irradiation of pure water ice and the temporal evolution of gases $H_2$, O, OH, $H_2O$, and $O_2$ escaping from it. Crystalline and amorphous ices of the same thickness were desorbed at 100 K. Overall number of desorption species and sputtering yield are higher in the case of amorphous ice compared to crystalline ice. This is easily understood based on larger voids in amorphous ice promoting easy sputtering, while crystalline ice is compact, ordered, and doesn't have sufficient mobility to sputter. At 2 keV, the averaged desorption per keV is approximately similar in both ices for $H_2$, O, OH, and $H_2O$, while $O_2$ the desorption is 1.6 times higher in amorphous ice. With 1 keV electron, the desorption of all the gases is higher in amorphous ice where $O_2$ is 2.2 times higher, indicating chemistry at shallow depths contributes significantly to the electron-induced sputtering yields of these species. Electron-induced surface ice chemistry followed by sputtering of volatiles could contribute significantly to permanent or transient exospheres of many solar system ice-covered bodies, including moons of outer planets, asteroids, comets, and permanently shadowed regions of the Moon. When not renewed through fresh ice depositions, these surface ice layers could be enriched in $H_2O_2$ over time. With increasing electron energy, the electron penetration depths increase, making subsurface ice chemistry predominant per keV of energy deposited into the ice at higher electron energies. Further work at higher electron energy, from 10 keV to 100 keV and higher would be needed to obtain quantitative sputtering yields by electrons from ice. Also, the influence of ice composition on sputtering yields of volatile species needs to be investigated in the future.

**Acknowledgment**

YNM gratefully acknowledges the Swedish Research Council for the financial support of Project No. 2018-06783. MSG thanks JPL for support through the JPL-Caltech JROC program. YLY acknowledges support from the Virtual Planetary Laboratory of the University of Washington and NSF grant to Caltech. This work is also supported by NASA SSW and HW grants to MSG and BLH. This work was carried out at the Jet Propulsion Laboratory, California Institute of Technology, and was under a contract with the National Aeronautics and Space Administration. Finally, we acknowledge Dr. Jiazheng Li at California Institute of Technology for his insights on $O_2$ sputtering.

**References**

Abdulgalil, A. G. M., Rosu-Finsen, A., Marchione, D., Thrower, J. D., Collings, M. P., McCoustra, M. R. S., 2017. Electron-Promoted Desorption from Water Ice Surfaces: Neutral Gas-Phase Products. ACS Earth and Space Chemistry. 1, 209-215.

Barnett, I. L., Lignell, A., Gudipati, M. S., 2012. Survival Depth of Organics in Ices under Low-energy Electron Radiation (≤2 keV). The Astrophysical Journal. 747, 13.

Berdis, J. R., Gudipati, M. S., Murphy, J. R., Chanover, N. J., 2020. Europa's surface water ice crystallinity: Discrepancy between observations and thermophysical and particle flux modeling. Icarus. 341, 113660.

Cassidy, T. A., et al., 2013. Magnetospheric ion sputtering and water ice grain size at Europa. Planetary and Space Science. 77, 64-73.

Cooper, J. F., Johnson, R. E., Mauk, B. H., Garrett, H. B., Gehrels, N., 2001. Energetic Ion and Electron Irradiation of the Icy Galilean Satellites. Icarus. 149, 133-159.

Davis, M. R., Meier, R. M., Cooper, J. F., Loeffler, M. J., 2021. The Contribution of Electrons to the Sputter-produced $O_2$ Exosphere on Europa. The Astrophysical Journal. 908, L53.

DeSimone, A. J., Orlando, T. M., 2015. $H_2O$ and $O(P-3(J))$ photodesorption from amorphous solid water deposited on a lunar mare basalt. Icarus. 255, 44-50.

Dohnalek, Z., Kimmel, G. A., Ayotte, P., Smith, R. S., Kay, B. D., 2003. The deposition angle-dependent density of amorphous solid water films. Journal of Chemical Physics. 118, 364-372.

Floyd, G. R., Prince, R. H., 1972. Production of Ionized Water Clusters by Electron Bombardment of Ice. Nature Physical Science. 240, 11-12.

Galli, A., et al., 2016. Surface charging of thick porous water ice layers relevant for ion sputtering experiments. Planetary and Space Science. 126, 63-71.

Galli, A., et al., 2018. 0.2 to 10 keV electrons interacting with water ice: Radiolysis, sputtering, and sublimation. Planetary and Space Science. 155, 91-98.

Galli, A., Vorburger, A., Wurz, P., Tulej, M., 2017. Sputtering of water ice films: A re-assessment with singly and doubly charged oxygen and argon ions, molecular oxygen, and electrons. Icarus. 291, 36-45.

Hall, D. T., Strobel, D. F., Feldman, P. D., McGrath, M. A., Weaver, H. A., 1995. Detection of an oxygen atmosphere on Jupiter's moon Europa. Nature. 373, 677-679.

Heide, H. G., 1984. Observations on ice layers. Ultramicroscopy. 14, 271-278.
Heritier, K. L., et al., 2018. On the origin of molecular oxygen in cometary comae. Nature Communications. 9.
Johnson, R. E., Leblanc, F., Yakshinskiy, B. V., Madey, T. E., 2002. Energy distributions for desorption of sodium and potassium from ice: The Na/K ratio at Europa. Icarus. 156, 136-142.
Johnson, R. E., Sundqvist, B. U. R., 2018. Sputtering and detection of large organic molecules from Europa. Icarus. 309, 338-344.
Kollmann, P., et al., 2018. Electron Acceleration to MeV Energies at Jupiter and Saturn. Journal of Geophysical Research: Space Physics. 123, 9110-9129.
Leonard, E. J., Pappalardo, R. T., Yin, A., 2018. Analysis of very-high-resolution Galileo images and implications for resurfacing mechanisms on Europa. Icarus. 312, 100-120.
Li, J., Gudipati, M. S., Mishra, Y. N., Liang, M.-C., Yung, Y. L., 2022. Oxidant generation in the ice under electron irradiation: Simulation and application to Europa. Icarus. 373, 114760.
Li, J., Gudipati, M. S., Yung, Y. L., 2020. The influence of Europa's plumes on its atmosphere and ionosphere. Icarus. 352, 113999.
Meier, R. M., Loeffler, M. J., 2020. Sputtering of water ice by keV electrons at 60 K. Surface Science. 691, 121509.
Neugebauer, M., 1990. Spacecraft observations of the interaction of active comets with the solar wind. Reviews of Geophysics. 28, 231-252.
Nordheim, T. A., Hand, K. P., Paranicas, C., 2018. Preservation of potential biosignatures in the shallow subsurface of Europa. Nature Astronomy. 2, 673-679.
Orlando, T. M., Sieger, M. T., 2003. The role of electron-stimulated production of O2 from water ice in the radiation processing of outer solar system surfaces. Surface Science. 528, 1-7.
Paige, D. A., et al., 2010. Diviner Lunar Radiometer Observations of Cold Traps in the Moon's South Polar Region. Science. 330, 479-482.
Paranicas, C., Cooper, J. F., Garrett, H. B., Johnson, R. E., Sturner, S. J., 2009 Europa's Radiation Environment and ITs Effects on the Surface. In: R. T. Pappalardo, W. B. McKinnon, K. Khurana, (Eds.), Europa. The University of Arizona Press, Tucson, Arizona, USA, pp. 529-544.
Paranicas, C., et al., 2007. Europa's near-surface radiation environment. Geophysical Research Letters. 34.
Petrik, N. G., Kavetsky, A. G., Kimmel, G. A., 2006. Electron-Stimulated Production of Molecular Oxygen in Amorphous Solid Water. The Journal of Physical Chemistry B. 110, 2723-2731.
R.E. Johnson, R. W. C., J.F. Cooper, C. Paranicas, M.H. Moore, M.C. Wong, 2004. Radiation Effects on the Surfaces of the Galilean Satellites. Cambridge Univ. Press, Cambridge.
Raponi, A., et al., 2018. Variations in the amount of water ice on Ceres' surface suggest a seasonal water cycle. Science Advances. 4, eaao3757.
Roth, L., et al., 2016. Europa's far ultraviolet oxygen aurora from a comprehensive set of HST observations. Journal of Geophysical Research: Space Physics. 121, 2143-2170.

Shi, M., Baragiola, R., Grosjean, D., Johnson, R., Jurac, S., Schou, J., 1995. Sputtering of water ice surfaces and the production of extended neutral atmospheres. Journal of Geophysical Research: Planets. 100, 26387-26395.

Sieger, M. T., Simpson, W. C., Orlando, T. M., 1998. Production of O2 on icy satellites by electronic excitation of low-temperature water ice. Nature. 394, 554-556.

Smyth, W. H., Marconi, M. L., 2006. Europa's atmosphere, gas tori, and magnetospheric implications. Icarus. 181, 510-526.

Standards, N. I. o., Technology, 2000. NIST Chemistry Webbook: NIST Standard Reference Database Number 69. NIST.

Teolis, B. D., et al., 2010. Cassini Finds an Oxygen–Carbon Dioxide Atmosphere at Saturn's Icy Moon Rhea. Science. 330, 1813.

Teolis, B. D., Plainaki, C., Cassidy, T. A., Raut, U., 2017. Water Ice Radiolytic O2, H2, and H2O2 Yields for Any Projectile Species, Energy, or Temperature: A Model for Icy Astrophysical Bodies. Journal of Geophysical Research: Planets. 122, 1996-2012.

Teolis, B. D., Waite, J. H., 2016. Dione and Rhea seasonal exospheres revealed by Cassini CAPS and INMS. Icarus. 272, 277-289.

Weiss, B. P., et al., 2021. Searching for Subsurface Oceans on the Moons of Uranus Using Magnetic Induction. Geophysical Research Letters. 48, e2021GL094758.

Yao, Y., Giapis, K. P., 2018. Reply to "On the origin of molecular oxygen in cometary comae". Nature Communications. 9, 2581.

Zheng, W., Jewitt, D., Kaiser, R. I., 2006a. Formation of Hydrogen, Oxygen, and Hydrogen Peroxide in Electron-irradiated Crystalline Water Ice. The Astrophysical Journal. 639, 534-548.

Zheng, W., Jewitt, D., Kaiser, R. I., 2006b. Temperature Dependence of the Formation of Hydrogen, Oxygen, and Hydrogen Peroxide in Electron-Irradiated Crystalline Water Ice. The Astrophysical Journal. 648, 753-761.

Zheng, W., Jewitt, D., Kaiser, R. I., 2007. Electron irradiation of crystalline and amorphous D2O ice. Chemical Physics Letters. 435, 289-294.

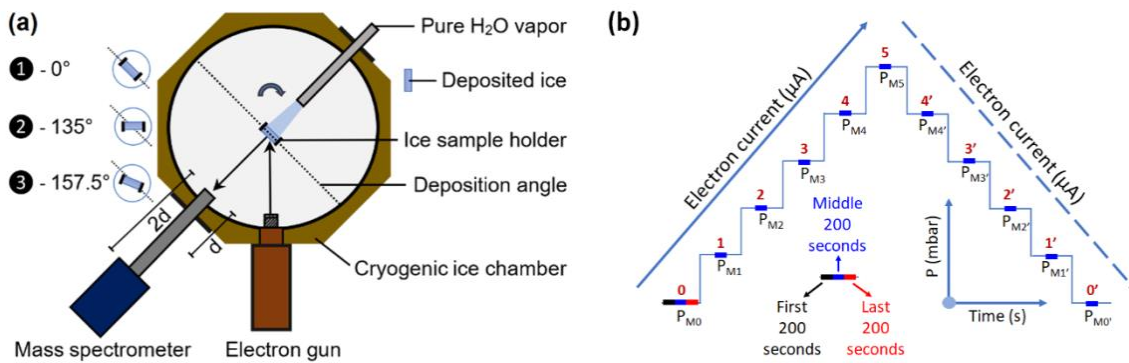

**Figure 1.** (a) The top view of the experimental setup showing the cryogenic ice chamber, ice sample, electron gun, and mass spectrometer. Note that the deposited ice symbols are indicated outside of the chamber here for a better illustration. The ice deposition is performed at position 1 at 0°, and the electron-irradiation study is conducted at position 3 at 157.5° to the deposition angle and quadrupole distance 2d. (b) Illustration of Partial pressure (P) vs. Time (s) spectra recorded by the mass spectrometer at electron currents 0-5 µA (upwards) and 5-0 µA (downwards). The "ladder-shaped" spectra are observed at 2 keV for all the gases (see Figures 2 and 3). The time of each step of the ladder is 600 seconds and each step is divided into 3 parts with the first 200 seconds (black) middle 200 seconds (blue) and last 200 seconds (red).

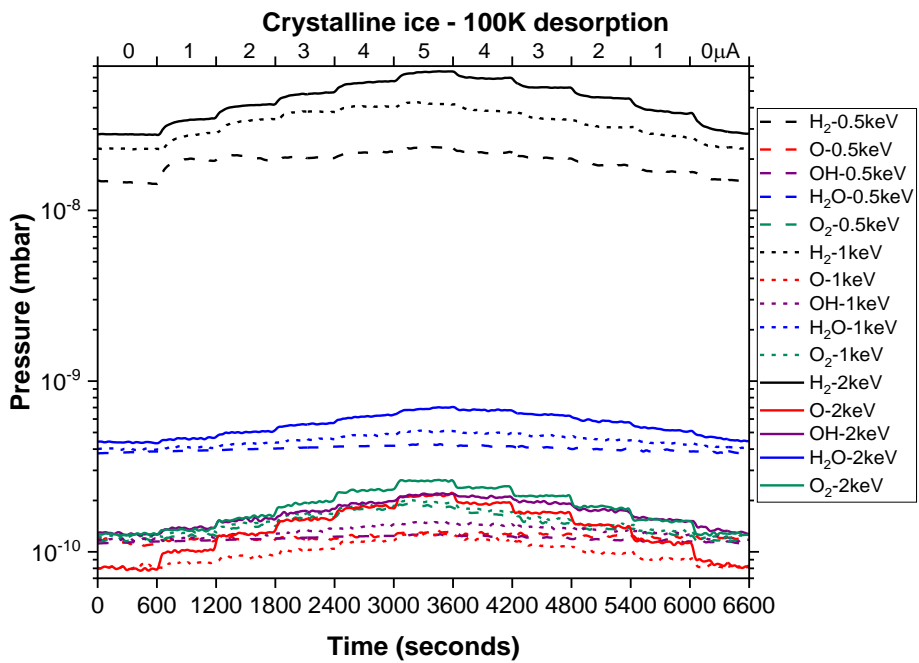

**Figure 2.** P vs. t spectra of gases desorbed at 100 K from electron-irradiated crystalline ice at energies of 0.5, 1, and 2 keV. The electron currents (I) are varied from 0-5-0 µA with a step size of ~ 600 seconds of constant irradiation. A "ladder-wise" change in the partial pressure as the function of current is shown for the probed species at 2 keV.

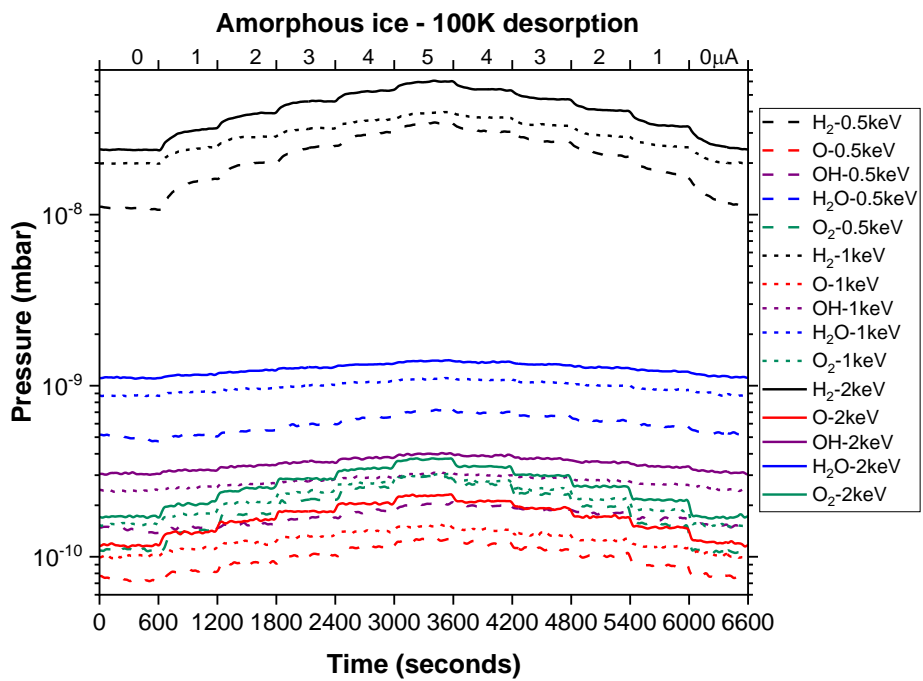

**Figure 3.** P vs. t spectra of gases desorbed at 100 K from the electron irradiated amorphous ice at energies 0.5, 1, and 2 keV. The electron currents (I) are varied from 0-5-0 µA with a step size of ~ 600 seconds of constant irradiation. A clear "ladder-wise" change in the pressure as the function of current can be seen for most of the probed species except for a relatively weaker change in $H_2O$, OH at 2 keV.

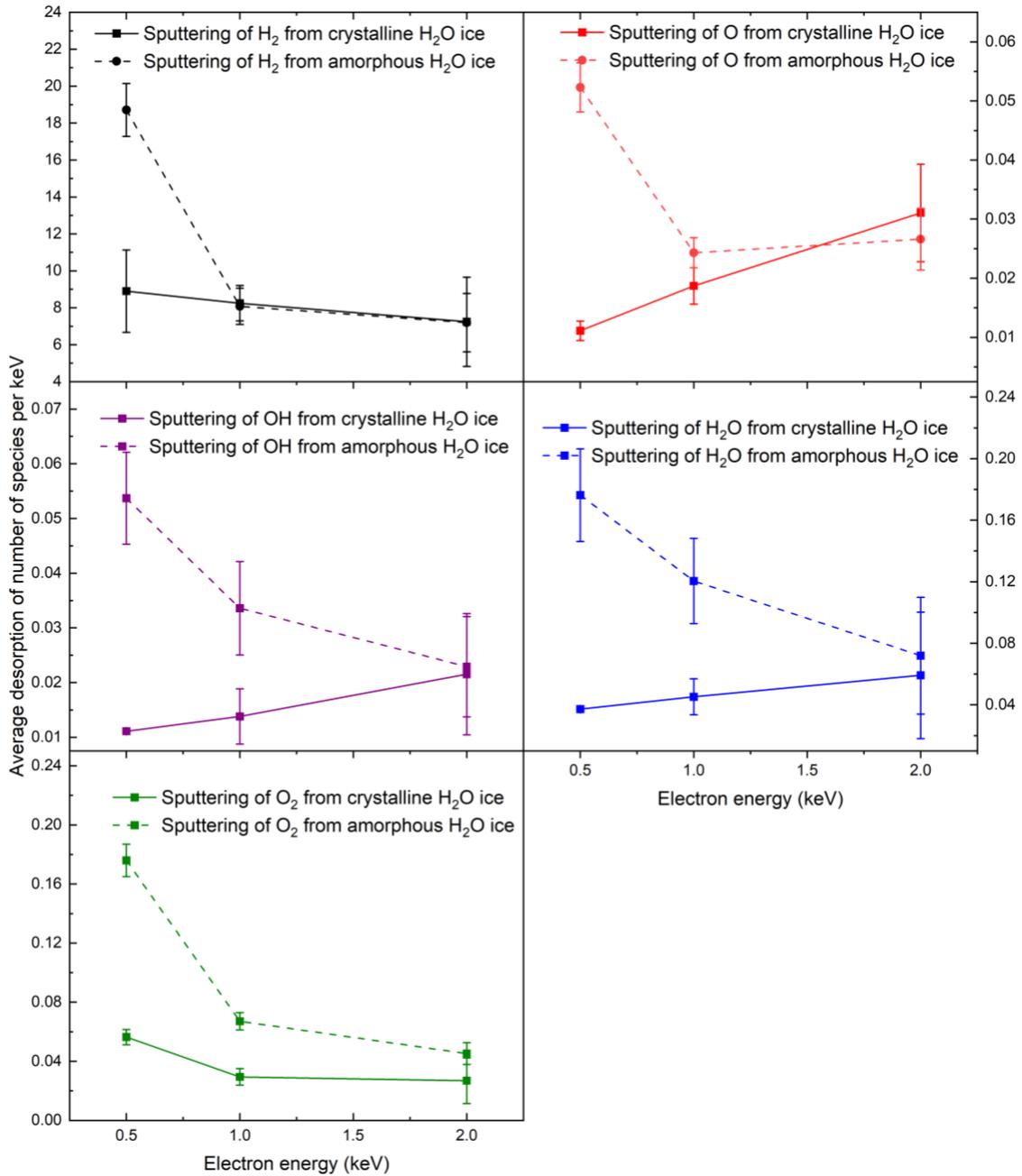

Figure 4: Dependence of the average number of desorbed species (normalized to the electron energy) on electron energy for amorphous and crystalline ice for sputtering of $H_2$, O, OH, $H_2O$, and $O_2$. For amorphous ice desorption, all the species decrease with electron energy.

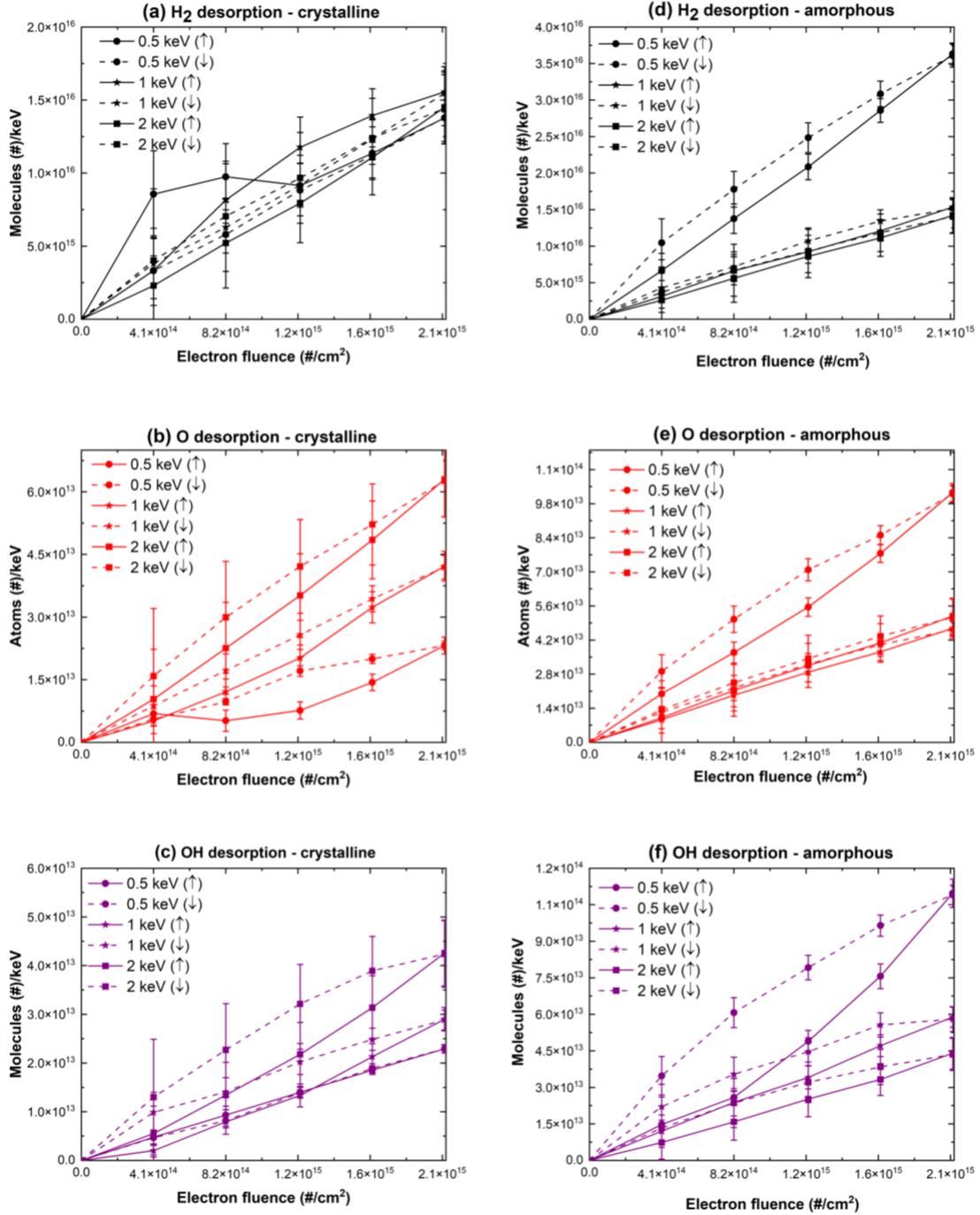

Figure 5: Uncorrected desorption of gases as the number of molecules-atoms cm$^{-2}$ per keV against the electron fluence (electrons cm$^{-2}$) from the crystalline and amorphous ice, respectively. For $H_2$ in (a, d), O atoms in (b, e), and OH in (c, f).

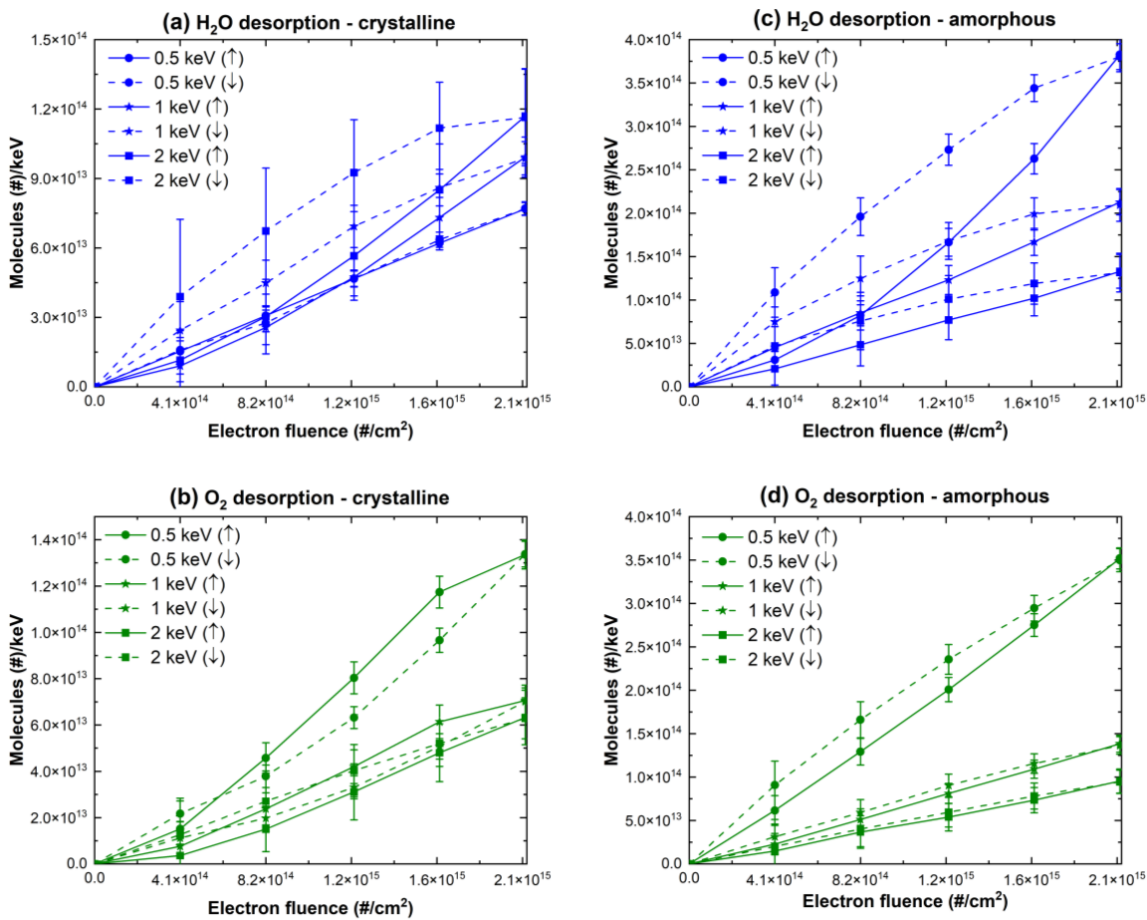

Figure 6: Uncorrected desorption of gases as the number of molecules-atoms cm$^{-2}$ per keV against the electron fluence (electrons cm$^{-2}$) from the crystalline and amorphous ice, respectively. For $H_2O$ in (a, c), and $O_2$ in (b, d).

# Appendix A1

*Optimization of the experimental configuration*

Figures S1(a) and S1(b) show the analog QMS spectra at 100 K from the electron-irradiated amorphous ice before and after the optimization of the experimental geometry in terms of QMS and sample configurations between positions 2 and 3 (see Figure 1(a)), respectively. In Figure S1(a), a plateau in the partial pressure of the entire mass range beyond 8-10 m/z was seen with position 2 and distance d configuration when irradiated with 2 keV electrons at a source current of 3 µA (red curve). At this time, when the current was reduced to 0.5 µA, the QMS signal changed from plateau to reach negative values after about 13 minutes, indicating significant changes in the electrostatics of the QMS taking long time to recover. The analog spectrum started to recover only after about 19 minutes of switching the current to 0.5 µA. The continuum plateau again as soon as the electron current was increased back to 3 µA. Changing the ice position to 180° did not improve the situation. We think this peculiar behavior of the QMS was due to the direct viewing of the electron-gun outlet by the QMS ion-region, causing significant interference by the active electron gun on the electrostatics of the QMS. To mitigate this strong interference by the electron gun in operation on the QMS, we changed the configuration to the position 3, whereby the QMS was pulled away from the main vacuum chamber into a side arm with an extension nipple. Subsequently, we optimized the ice angle between 135° and 180° to the ice-deposition reference position. We found that optimal angle was when the surface of the ice was equiangular with the electron gun and the QMS arm, at 157.5° from the initial position. With this new configuration we did not notice any interference by the electron gun to the QMS signal with electrons of the highest energy and flux used in our studies, namely, 2 keV and 5 µA. Figure S1(b) shows the analog mass spectra of the ice under this configuration for 2 keV and 0, 3, 5 µA. The usual analog mass spectra, with peaks of $H_2$, O, OH, $H_2O$, $O_2$, $N_2$, etc., can now clearly be seen. This kind of peculiar behaviors was not observed from earlier studies, therefore, it is important to document such a strong interference by an electron source on the QMS signals, when these two instruments are mounted in proximity. In such scenarios, conducting experiment where masses of various species are tracked against irradiation fluence (time) without prior verification whether the mass spectrometer is functioning properly or not, would lead to recording wrong data and wrong interpretations. Further, as an extra precaution, we followed maxima and minima of different species to ensure that at no point electrostatic interference occurred, in which case those two mass counts would be much closer (no mass spectrum, but electrostatic continuum). For example, to follow sputtering of O atoms, we were monitoring m/z 15 and 16. For $H_2O$ we monitored m/z 18 and (19 or 20).

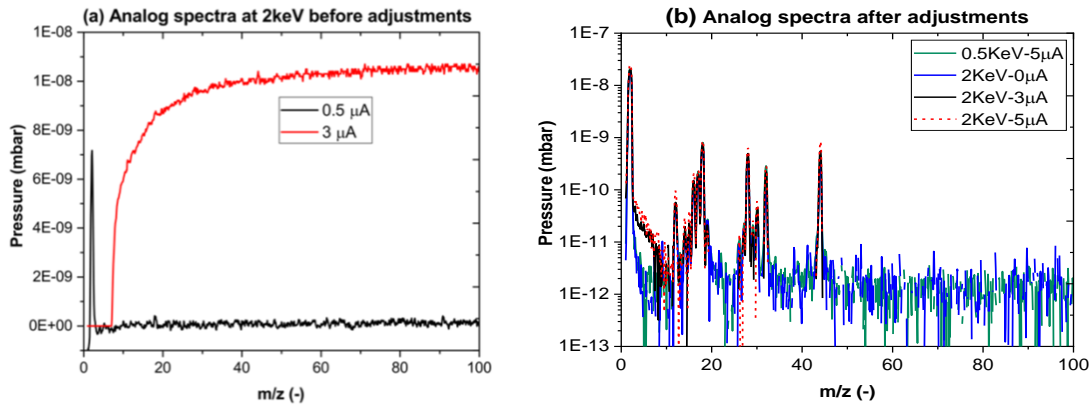

Figure S1. Analog mass spectra of amorphous ices (100 K) irradiated by electrons (0.5-2 keV) at position 2-135° to the deposition angle (pos. 1-0°) in (a) and at position 3-157.5° to the deposition angle in (b). The deposition angles are indicated in Figure 1(a). In (a) before adjustments in the deposition angle and quadrupole distance d, charge-induced anomalies in the mass spectra as continuum are seen. In (b) after the adjustments in deposition angle and extending the quadrupole distance from d to 2d, the corrected normal analog mass spectra are observed.

# Appendix A2

*Determination of mean pressure*

In P vs. t plots, the raw mass spectrometer data ($P_O$) for each gas was represented by n (rows) ×1 (column) matrix. The background pressure ($P_{BG}$) matrix for the corresponding gas is calculated from the final pressure ($P_F$) and initial pressure ($P_I$) values by extrapolating from initial pressure value ($P_I$) to the final pressure value ($P_F$). Therefore, the background pressure matrix was also represented by n (rows) ×1 (column) matrix. Note that the $P_F$ and $P_I$ are the values of partial pressure (see Figure 2 and 3) recorded during the first and last 10 minutes of P vs. t study keeping electron gun OFF (current = 0 µA). Therefore, the actual partial pressure of the gas ($P_R$) is deduced from column-wise subtraction ($P_O$-$P_{BG}$) which corresponds to the desorption pressure of the species leaving the electron-irradiated ice. From the $P_R$ plots the mean pressure ($P_M$) values are deduced for I = 0-5 µA and I' = 5'-0' µA. It is indicated as a solid blue line in the "ladder curve" illustration in Figure 1(b). Here, $P_M$ is extracted from the averaged $P_R$ of middle 200 seconds of each electron current exposure for a duration of 600 seconds. This process generated a total of 11 $P_M$ values such as $P_{M0}$, $P_{M1}$, $P_{M2}$, ····· $P_{M5}$ for upward step and similarly from $P_{M4'}$, $P_{M3'}$, ····, $P_{M0'}$ for the downward step. The subscript 0, 0', 1, 1', 2, 2' and so on indicate the electron current values for upward and downward steps, respectively. Similarly, mean background pressure is calculated for the associated time as $P_{BG0}$, $P_{BG1}$…$P_{BG5}$, $P_{BG5'}$…$P_{BG1'}$ and $P_{BG0'}$. Now, to deduce ΔP per µA, the subtraction is operated between two values of $P_M$ as ($P_{M5}$ – $P_{MBG5}$), ($P_{M4}$ – $P_{MBG4}$), ········, ($P_{M1}$ – $P_{MBG1}$) – upward step. Similarly, ΔP values are deduced from ($P_{M5'}$ – $P_{MBG5'}$), ($P_{M4'}$ – $P_{MBG4'}$), ········, ($P_{M1'}$ – $P_{MBG1'}$) for the downward step. Finally, the effective averaged change in partial pressure, $\Delta P_e$ for each current value as two matrices (one for upward and another for downward) steps are calculated from the summation of each resulting ΔP values.

# Appendix A3

*Pressure correction*

The procedure to deduce the correct desorption can be expressed as following:

$O_{(corrected)}$ = O – (0.029*$H_2O$) – (0.1*$O_2$) because O is generated from both $H_2O$ and $O_2$. Similarly, $OH_{(corrected)}$ = OH – (0.27*$H_2O$). The total (integrated) $H_2O$ was calculated by adding $H_2O$, OH, and O peaks, such that: $H_2O_{(corrected)}$ = $H_2O$ + (0.029*$H_2O$) + (0.27*$H_2O$). Finally, $O_{2(corrected)}$ = $O_2$ + ((0.1*$O_2$)/2) because 1 atom of O is generated from ½ $O_2$ molecules.